\documentclass[%
twocolumn,
superscriptaddress,
 amsmath,amssymb,
 aps,
pra,
]{revtex4-2}

\usepackage{graphicx}
\usepackage{dcolumn}
\usepackage{bm}
\usepackage{color}
\usepackage{ulem}
\usepackage{stackengine}
\usepackage{braket}
\usepackage{multirow}

\newcommand{\figref}[1]{Fig.~\ref{fig:#1}}
\newcommand{\secref}[1]{Sec.~\ref{sec:#1}}
\newcommand{\equref}[1]{Eq.~(\ref{eq:#1})}
\newcommand{\tabref}[1]{Table~\ref{tab:#1}}

\begin{document}

\title{
Universal Efimov spectra and fermionic doublets in highly mass-imbalanced cold-atom mixtures with van der Waals and dipole interactions
}
\author{Kazuki Oi}
\email{kazuki.oi.t8@dc.tohoku.ac.jp}
\affiliation{Department of Physics, Tohoku University, Sendai 980-8578, Japan}
\affiliation{Department of Engineering Science, University of Electro-Communications, Chofu, Tokyo 182-8585, Japan}
\author{Shimpei Endo}
\email{shimpei.endo@uec.ac.jp}
\affiliation{Department of Engineering Science, University of Electro-Communications, Chofu, Tokyo 182-8585, Japan}

\date{\today}

\begin{abstract}
We study the Efimov states in highly mass-imbalanced three-body systems composed of two identical heavy atoms and one light atom, focusing on the Er-Er-Li and Dy-Dy-Li cold-atom mixtures with strong dipole-dipole interactions between the heavy atoms. By solving the Born-Oppenheimer equation for varying $s$-wave scattering lengths between the heavy and light atoms, we demonstrate for both bosonic and fermionic systems that the Efimov spectra and hence the three-body parameters are universal even with the dipole interaction comparable in strength to the van der Waals interaction. While the bosonic systems exhibit Efimov states only in the $M_z=0$ channel, the fermionic systems show a characteristic doublet of the Efimov states in the $M_z=0$ and $M_z = \pm 1$ channels due to the interplay of finite angular momentum and the anisotropy of the dipole interaction. Both numerical results and analytical formulae obtained with the first-order perturbation show that the ratio of the three-body parameters between these two fermionic channels
exhibits universality, particularly well in the limit of large mass imbalance. Leveraging this universality, we provide quantitative predictions for the values and ratios of the three-body parameters for experimentally relevant Er-Li and Dy-Li isotopes.
\end{abstract}

\maketitle

\section{Introduction}
Physical systems are regarded to be universal when they display the same behaviors and can be described with a unified effective theory. A prominent example of universal few-body quantum phenomena is the Efimov effect~\cite{efimov1970energy,efimov1973energy,naidon2017efimov,RevModPhys.89.035006,d2018few,Braaten2006259}, where an infinite series of three-body bound states can emerge universally when the inter-particle $s$-wave scattering length is large. Systems with large scattering lengths arise in various contexts, such as $^4$He atoms~\cite{kunitski2015observation,PhysRevLett.38.341}, halo nuclei~\cite{AnnRev_HamPlatt,PhysRevLett.120.052502,hammer2017effective,hoyle1954nuclear}, and cold atoms~\cite{kraemer2006evidence}. To uncover the nature of Efimov states in these systems, extensive studies have been conducted both theoretically and experimentally, particularly in cold atoms where the $s$-wave scattering length can be precisely controlled via Feshbach resonances~\cite{inouye1998observation,chin2010feshbach}. A notable finding is that the key parameter characterizing the Efimov states, the three-body parameter $a_-$  ($s$-wave scattering length at which the Efimov states dissociate into three atoms), is universally determined by the van der Waals length scale~\cite{PhysRevLett.107.120401,gross2011study}. This suggests that short-range details of the atoms, such as their electronic structures and spin states, are irrelevant, and physical properties of the Efimov states are solely determined by the long-range features of the atoms. This universality has been demonstrated for various atomic species, including identical bosons~\cite{PhysRevLett.107.120401,gross2011study,pascaleno3BP1,pascaleno3BP2,PhysRevLett.108.263001}, and bosonic mixtures~\cite{PhysRevLett.109.243201,PhysRevA.95.062708}, except for systems close to a narrow Feshbach resonance~\cite{PhysRevLett.111.053202,PhysRevLett.123.233402,PhysRevLett.125.243401,johansen2017testing,schmidt2012efimov,PhysRevA.103.052805,PhysRevLett.93.143201,PhysRevLett.100.140404,PhysRevA.86.052516,PhysRevA.107.023301,PhysRevLett.132.133402}.

While the understanding of the Efimov states has advanced considerably for the bosonic systems, less has been explored for systems involving identical fermions. In fermionic systems, the centrifugal barrier originating from the non-zero angular momentum due to the antisymmetrization suppresses the formation of Efimov states for atoms with similar masses, which accounts for the stability of the unitary Fermi gases~\cite{PhysRevA.67.010703,PhysRevLett.93.090404}. To overcome this, a mass-imbalanced mixture with a mass ratio exceeding 13.6 is required~\cite{efimov1973energy,PhysRevA.67.010703}. This condition is satisfied in cold-atom mixtures such as Yb-Li~\cite{PhysRevLett.106.205304,PhysRevA.96.032711,Dowd_2015,PhysRevX.10.031037}, Er–Li~\cite{ErLiFR1,ErLiFR2,kalia2025creationdegeneratebosebosemixture} and Dy–Li~\cite{DyLifeshbach}. In particular, broad Feshbach resonances have been observed for Er–Li~\cite{ErLiFR1,ErLiFR2} and Dy–Li~\cite{DyLifeshbach}, offering promising platforms for the near-future observation of the fermionic Efimov states of angular momentum $L=1$. However, Er and Dy atoms possess large magnetic dipole moments~\cite{Chomaz_2023}, making the dipole-dipole interaction comparable in strength to the van der Waals interaction. Due to the anisotropy of the dipole interaction, the total angular momentum $L$ is no longer a good quantum number, whereas its azimuthal value $M_z$ along the external magnetic field axis remains conserved~\cite{PhysRevLett.106.233201,PhysRevLett.107.233201,OiEndo2024}. This leads to a fundamental question: ``Does the universality of the Efimov states persist even in the presence of strong dipole interactions? How do the fermionic Efimov states differ from the bosonic ones?"

These questions were partially addressed in a previous study~\cite{OiEndo2024}, where the $M_z=0$ state was studied and the universality of the Efimov states have been demonstrated at the unitary limit of the heavy-light scattering length. Here, we extend the analysis to the entire values of the heavy-light $s$-wave scattering length, and investigate the universality across the full Efimov spectrum. In the bosonic system, the Efimov states appear only in the $M_z=0$ channel for the mass ratios of our interests, Er-Li and Dy-Li. We show that the Efimov spectrum and hence the three-body parameter $a_-$ is universal. In contrast, in the fermionic system, the Efimov states emerge in both the $M_z=0$ and $M_z=\pm 1$ states. Due to the interplay of finite angular momenta and the anisotropy of the dipole interaction, their degeneracy is lifted, analogous to the splitting of a $p$-wave Feshbach resonance~\cite{PhysRevA.69.042712,PhysRevA.70.030702,PhysRevA.71.045601,PhysRevA.88.012710,PhysRevA.85.051602,PhysRevA.100.050701,luciuk2016evidence,peng2025precision}. This leads to a characteristic doublet structure in the Efimov spectrum. We show that the ratio of the three-body parameters $a_-$ between these two channels is universal,  especially in the limit of large mass imbalance. These results provide a practical experimental criterion for identifying Efimov states in dipolar systems: one three-body loss peak per limit cycle for the bosons, and two peaks per cycle for the fermions. We provide quantitative predictions for the values and ratios of the three-body parameters for experimentally relevant Er–Li and Dy–Li isotopes, paving the way for exploring universal fermionic Efimov physics.

This paper is organized as follows: in \secref{model}, we introduce the model and method we use in our calculations. In \secref{result}, the results of our numerical calculations presented and compared with our perturbative analysis, followed by our estimates of the three-body parameters for Er-Li and Dy-Li isotopes. We conclude in \secref{Concl}. Throughout the paper, the natural unit $\hbar=1$ is used.

\section{\label{sec:model}Model \& Method}
We follow the method of Ref.~\cite{OiEndo2024}; we consider a three-body system consisting of two heavy particles (mass $M$) and a light particle (mass $m$) without any internal states, which models spin-polarized cold atoms. The heavy particles, assumed to be either identical bosonic or fermionic atoms, interact via both the van der Waals interaction and dipole interaction. The light and heavy particles are assumed to interact via the zero-range contact interaction with an $s$-wave scattering length $a^{(\rm{HL})}$ whose value can be controlled by the Feshbach resonance~\cite{inouye1998observation,chin2010feshbach}. While the zero-range approximation is well-suited for describing the Efimov states near broad Feshbach resonances, it becomes inadequate for narrow resonances~\cite{PhysRevLett.111.053202,PhysRevLett.123.233402,PhysRevLett.125.243401,johansen2017testing,schmidt2012efimov,PhysRevA.103.052805,PhysRevLett.93.143201,PhysRevLett.100.140404,PhysRevA.86.052516,PhysRevA.107.023301,PhysRevLett.132.133402}, where it is necessary to incorporate the closed-channel degree of freedom.

In highly mass-imbalanced systems, such as Er-Li~\cite{ErLiFR1,ErLiFR2} or Dy-Li~\cite{SoaveGrimm2022} cold-atom mixtures, the Born-Oppenheimer approximation gives good quantitative descriptions for the Efimov physics~\cite{efimov1973energy,FONSECA1979273,PhysRevA.67.010703,PhysRevLett.109.243201}. The Schr\"odinger equation of the relative motion between the heavy particles is analytically written with the potential induced by the light particle $V_{\mathrm{BO}}(r)$ as follows~\cite{FONSECA1979273,OiEndo2024,naidon2017efimov}:
\begin{equation}
\label{eq:BOvdwDipoleSchrodinger}
   \left[-\frac{\nabla_{\bm{r}}^2}{M} +V_{\mathrm{BO}}(r)-\frac{C_6}{r^6}+\frac{C_{\mathrm{dd}}(1-3\cos^2\theta)}{r^3}\right]\psi(\bm{r})=E\psi(\bm{r}),
\end{equation}
where $\theta$ is an angle measured from the $z$ axis taken to be parallel to the orientation of the dipoles (parallel to the external magnetic field). $C_6$ and $C_{\mathrm{dd}}$ are the coefficients of the van der Waals and dipole interactions between the heavy particles, from which we can define their characteristic length scales: the van der Waals length $r_{\mathrm{vdw}}\equiv\frac{1}{2}(MC_6)^{1/4}$ and dipole length $a_{\mathrm{dd}}\equiv MC_{\mathrm{dd}}/3$. $V_{\mathrm{BO}}(r)$ is given as~\cite{FONSECA1979273}

\begin{equation}
    V_{\mathrm{BO}}(r)\equiv
    -\dfrac{1}{2mr^2}\left(\dfrac{r}{a^{(\rm{HL})}}+W\left(e^{-\frac{r}{a^{(\rm{HL})}}}\right)\right)^2 \theta\left(\frac{r}{a^{(\rm{HL})}}+1\right) ,
\end{equation}
where $W$ is the Lambert $W$ function. $\theta(x)$ is the Heaviside step function introduced to get a smooth Born-Oppenheimer potential $V_{\mathrm{BO}}$ at $r=|a^{(\rm{HL})}|$ when $a^{(\rm{HL})}$ is negative. 

In the short-range region $r\ll |a^{(\rm{HL})}|$, a boundary condition is needed to avoid the collapse induced by $V_{\mathrm{BO}} \propto -1/r^2$~\cite{efimov1970energy,Braaten2006259,naidon2017efimov} and by the van der Waals and dipole potentials~\cite{PhysRevA.58.1728,PhysRevA.64.010701,PhysRevA.78.012702,OiEndo2024}. In our analysis, we introduced the hard-wall boundary condition $\psi(|\bm{r}|=R_{\rm{min}})=0$ at a short distance $R_{\rm{min}}\lesssim r_{\rm{vdw}}$, which is equivalent to imposing the quantum defect $K^c$ as
\begin{equation}
\label{eq:QDT_uKc_definition}u_\ell(r ) = f^c (r) - K^c g^c(r) 
\end{equation}
between the two independent solutions $f^c $ and $g^c $ at short distance. Indeed, the quantum defect parameter $K^c$ is related with $R_{\rm{min}}$ as~\cite{OiEndo2024}
\begin{align}
K^c &= - \tanh\left(\frac{\pi|s_\ell|}{4}\right)\frac{\mathrm{Re}\left[J_{\frac{s_\ell}{2}}\left(\frac{2r_{\rm vdw}^2}{R^2_{\mathrm{min}}} \right)\right]}{\mathrm{Im}\left[J_{\frac{s_\ell}{2}}\left(\frac{2r_{\rm vdw}^2}{R^2_{\mathrm{min}}}\right)\right]}\label{eq:KcRminrelation1}\\
& \simeq  - \frac{1}{\tan\left[\frac{2r_{\rm vdw}^2}{R^2_{\mathrm{min}}} - \frac{\pi}{4} \right]}\label{eq:KcRminrelation2},
\end{align}
where the asymptotic form of the Bessel function $J_n$ is used in the second line.

Due to the dipole interaction in \equref{BOvdwDipoleSchrodinger}, the total orbital angular momentum $L$, i.e. the angular momentum between the heavy particles, is no longer a good quantum number, while its projection along the $z$ axis $M_z$ is. The quantum statistics of the heavy particles plays a crucial role in the Efimov effect: for bosonic heavy particles, Efimov states can emerge only in the $M_z=0$ state for mass ratios $M/m \le 38.6...$~\cite{efimov1973energy,naidon2017efimov,kartavtsev2007low,endo2011universal} a condition relevant for Er-Li~\cite{ErLiFR1,ErLiFR2} or Dy-Li~\cite{SoaveGrimm2022} systems. Indeed, for a $M_z=2$ state for example, none of the orbital angular momentum channels $L=2,4,6...$ contributing to this state show the Efimov effect. In contrast, for fermionic heavy particles, Efimov states can appear in both the $M_z=0$ and $M_z=\pm 1$ states. In the absence of the dipole interaction, all these states are degenerate. As we demonstrate in \secref{result}, the dipole interaction lifts this degeneracy, while $M_z= +1$ and $M_z= -1$ states remain degenerate due to the azimuthal symmetry. Efimov states do not appear for $M_z =\pm 2$ (and similarly for larger $|M_z|$ states) because the minimum allowed angular momentum under the antisymmetrization, $L=3$, requires $M/m \ge 75.9...$ for the Efimov effect to occur~\cite{efimov1973energy,naidon2017efimov,kartavtsev2007low,endo2011universal}. Therefore, to investigate Efimov physics in Er-Li~\cite{ErLiFR1,ErLiFR2} or Dy-Li~\cite{SoaveGrimm2022} systems, we focus on $M_z= 0$ for bosons, and $M_z= 0$ and $M_z= \pm 1$ for fermions, which amounts to numerically solving the coupled-channel equation with $L=0,2,4,...$ and $L=1,3,5,...$, respectively.
\section{\label{sec:result}Results}
In this section, we show the results of our numerical calculations; we diagonalize \equref{BOvdwDipoleSchrodinger} by discretizing the radial coordinate in a non-uniform manner such that the Wentzel-Kramers-Brillouin (WKB) phase increment between adjacent grid points remains nearly constant. We typically take $R_{\rm{}max}= 400 r_{\mathrm{vdw}}- 12000r_{\mathrm{vdw}}$ with 3000$-$5000 grid points with the maximum angular momentum $\ell_{\rm{}max} =8-9 $. With these parameters, the numerical error of three-body binding energy and $a_-$ is confirmed to be less than 0.5\% in most cases, except near avoided crossings, where strong admixture of higher angular momentum channels increases the error to up to 4\%. Except for \secref{prediction} where the results of Er-Li and Dy-Li systems are quantitatively compared, we primarily present the results for Er-Li as the overall behavior of the Dy–Li system has been confirmed to be qualitatively similar. That is, in the bosonic system, we have chosen $M/m=27.5855\dots$ which corresponds to $^{166}$Er-$^6$Li and a scaling factor $e^{\frac{\pi}{|s_0|}}=4.64326\dots$. In the fermionic system, we have chosen $M/m=27.7521\dots$ which corresponds to $^{167}$Er-$^6$Li and $ e^{\frac{\pi}{|s_1|}}=8.26236\dots$.

\begin{figure}[!t]
\centering
\includegraphics[width=1.0\linewidth]{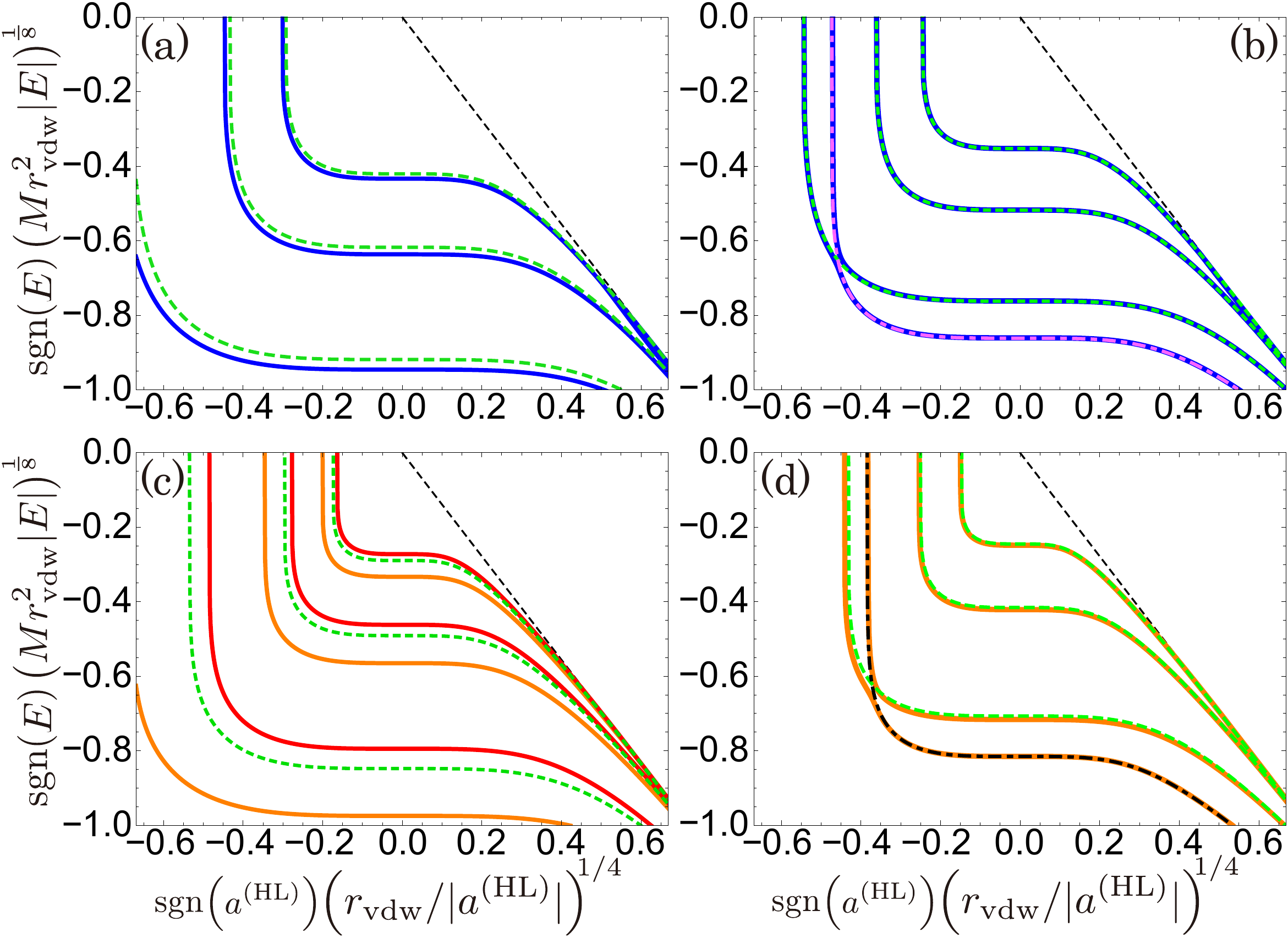} 
\caption{Binding energy of the Efimov trimer as a function of the heavy-light $s$-wave scattering length $a^{(\mathrm{HL})}$ for (a)(b) the bosonic ($M_z=0$) and (c)(d) fermionic ($M_z=0,\pm1$) Efimov states. (a) Results with (without) the dipole interaction are denoted by the blue solid (green dashed) curves for $a_{\mathrm{dd}}=0.86755... r_{\mathrm{vdw}}$, corresponding to $^{166}$Er-$^6$Li. (c) Results with the dipole interaction in the $M_z=\pm 1$ ($M_z=0$) states are shown by the upper red (lower orange) solid curves for 
$a_{\mathrm{dd}}=0.87054... r_{\mathrm{vdw}}$, corresponding to $^{167}$Er-$^6$Li. (b)(d) Cases showing avoided crossings with the higher partial-wave channels (a smaller dipole strength $a_{\mathrm{dd}}=0.1 r_{\mathrm{vdw}}$ is taken in (b)(d)). The pink [black] dashed-dotted curves in (b) [in (d)] represent the energy of the $d$-wave [$f$-wave] trimer, evaluated by including only the diagonal dipole contribution (see text). In (d), only the $M_z=0$ state is shown for clarity of the figure. The heavy–light dimer energy is indicated by black dashed lines. $R_{\mathrm{min}}$ is taken as follows: (a)
$R_{\mathrm{min}}=0.272 r_{\mathrm{vdw}}$, (b) $0.2641 r_{\mathrm{vdw}}$, (b) $0.27 r_{\mathrm{vdw}}$, (d) $0.2597 r_{\mathrm{vdw}}$.}
\label{fig:aHL_Energy}
\end{figure}

\subsection{\label{sec:EfimovSpectrum}Efimov spectrum}
We first show the Efimov spectrum as a function of the heavy-light scattering length $a^{(\rm{HL})}$. Figure~\ref{fig:aHL_Energy}~(a) shows the three-body binding energy for the bosonic $M_z=0$ state, obtained by numerically solving \equref{BOvdwDipoleSchrodinger} with a fixed $R_{\rm{min}}$. For $a_{\mathrm{dd}}=0.86755... r_{\mathrm{vdw}}$ (blue solid), corresponding to $^{166}$Er-$^6$Li, the results exhibit a typical Efimov spectrum with the discrete scale invariance: at $|a^{(\mathrm{HL})}|\gg r_{\mathrm{vdw}}$ and $|E|\ll 1/Mr_{\mathrm{vdw}^2}$, the spectrum is invariant under the transformation $a^{(\mathrm{HL})}\rightarrow e^{\frac{\pi}{|s_0|}}a^{(\mathrm{HL})},E\rightarrow (e^{\frac{\pi}{|s_0|}})^{-2}E$ where $e^{\frac{\pi}{|s_0|}}=4.64326\dots$. For $a^{(\mathrm{HL})}<0$, the three-body system is in the Borromean regime where three-body bound states appear without a two-body bound state. In contrast, for $a^{(\mathrm{HL})}>0$, the three-body bound states lie below the heavy-light dimer threshold (black dashed line) and approach it as $a^{(\mathrm{HL})}$ decreases, eventually merging into the threshold and dissociating. The binding energy in the presence of the dipole interaction (blue solid) is larger than that without the dipole interaction (green dashed). This enhancement arises from the attractive nature of the dipole interaction in the bosonic $L=0$ channel, as demonstrated in the unitary limit using the second-order perturbation theory in Refs.~\cite{bohn2009quasi,OiEndo2024}.

The three-body bound states shown in \figref{aHL_Energy}~(a) can be identified as the ground, first-excited, and second-excited Efimov states. They form part of a series that continues to higher excited Efimov states, which are not shown in \figref{aHL_Energy} or in the subsequent sections for clarity. On the other hand, we also find deeply bound states with energies $|E|\gtrsim 50/Mr_{\mathrm{vdw}}^2 $ which deviate significantly from the discrete scale-invariant behavior and thus cannot be regarderd as the Efimov state. We therefore assign the trimer of $|E| \simeq 1/Mr_{\mathrm{vdw}}^2$ as the ground Efimov state.

Figure~\ref{fig:aHL_Energy}~(b) shows the results with different $R_{\mathrm{min}}$, optimally chosen to ensure the appearance of a weakly bound $d$-wave dominant bound state (pink dashed-dotted) in addition to the $s$-wave dominant states. While the $s$-wave–dominant trimers form an infinite Efimov series, only a single $d$-wave-dominant trimer appears in the spectrum because the Efimov effect does not occur in the $d$-wave channel for the bosonic Er-Li mass ratio $M/m=27.5855\dots$. Without the dipole interaction, the $s$-wave (green dashed) and $d$-wave (pink dashed-dotted) states remain uncoupled because the angular momentum is a good quantum number. Once the dipole interaction is introduced (blue solid), an avoided crossing between the $s$-wave and $d$-wave dominant states occurs. However, such an avoided crossing is relatively rare; we have found that a fine-tuning of the parameter $R_{\mathrm{min}}$ is necessary for the higher partial bound states to appear near zero energy (a similar result was found in Ref.~\cite{OiEndo2024} at the unitary limit). Therefore, in most realistic cold-atom experiments, we expect that such an avoided crossing rarely occurs, and the Efimov spectrum behaves as in \figref{aHL_Energy}~(a).

The fermionic system (\figref{aHL_Energy}~(c)(d)) is found to show similar Efimov spectra. One notable difference is that there are three Efimov states, corresponding to a $M_z=0$ state (orange bottom solid) and two-fold degenerate $M_z=\pm 1$ states (red upper solid). In the absence of the dipole interaction (green dashed), all these states of $L=1$ are degenerate. The dipole interaction breaks the rotational symmetry, so that this degeneracy is lifted; the $M_z=0$ state is shifted toward a tighter binding, while the $M_z=\pm 1$ states are shifted toward a shallower binding. This behavior can be understood from the difference in the diagonal element of the dipole interaction: it is attractive $\bra{M_z=0}V_{dd}\ket{M_z=0}=-12a_{\mathrm{dd}}/5Mr^3<0$ for $M_z=0$, while it is repulsive $\bra{M_z=\pm1}V_{dd}\ket{M_z=\pm1}=6a_{\mathrm{dd}}/5Mr^3>0$ for $M_z=\pm1$.

As shown in \figref{aHL_Energy}~(d), the fermionic Efimov states can also exhibit avoided crossings with a higher angular momentum state, similar to the bosonic system; when an $f$-wave state (black dashed-dotted) happens to appear at low energy -- obtained using the single-channel approximation of \equref{BOvdwDipoleSchrodinger} in the $L=3$ channel considering only the diagonal dipole contribution -- it can couple to the fermionic Efimov states of $L=1$ (green dashed) via the dipole interaction, leading to an avoided crossing. We note however that as in the bosonic system, a fine-tuning of $R_{\mathrm{min}}$ is necessary for the avoided crossing to appear, suggesting that the fermionic Efimov spectra most likely behave as in \figref{aHL_Energy}~(c).

Figure~\ref{fig:aHL_Energy} has an important implication for the experimental identification of the fermionic Efimov states in Er-Li and Dy-Li systems: the emergence of a doublet structure in observables. In cold atoms, Efimov states are typically observed via enhanced three-body loss rates near the dissociation point $a_- <0$, at which the Efimov trimers dissociate into three particles~\cite{kraemer2006evidence,PhysRevLett.112.250404,PhysRevLett.113.240402,PhysRevLett.115.043201,PhysRevLett.101.203202,PhysRevLett.103.130404,PhysRevLett.105.023201,PhysRevLett.107.120401,gross2011study,PhysRevLett.112.190401}. Figure~\ref{fig:aHL_Energy}~(c) suggests that the fermionic system should exhibit two loss peaks, each corresponding to $M_z=0$ and $M_z=\pm1$. As the latter is two-fold degenerate, the resulting loss spectrum should display an asymmetric doublet structure, repeating periodically in accordance with the discrete scale invariance of Efimov physics. This behavior contrasts sharply with that of bosonic Efimov states~\cite{kraemer2006evidence,PhysRevLett.112.250404,PhysRevLett.113.240402,PhysRevLett.115.043201,PhysRevLett.107.120401,gross2011study,PhysRevLett.112.190401} and of distinguishable fermions~\cite{PhysRevLett.101.203202,PhysRevLett.103.130404,PhysRevLett.105.023201}, which produce a periodic loss pattern with a single peak per cycle.

\begin{figure}[!t]
\centering
\includegraphics[width=1.0\linewidth]{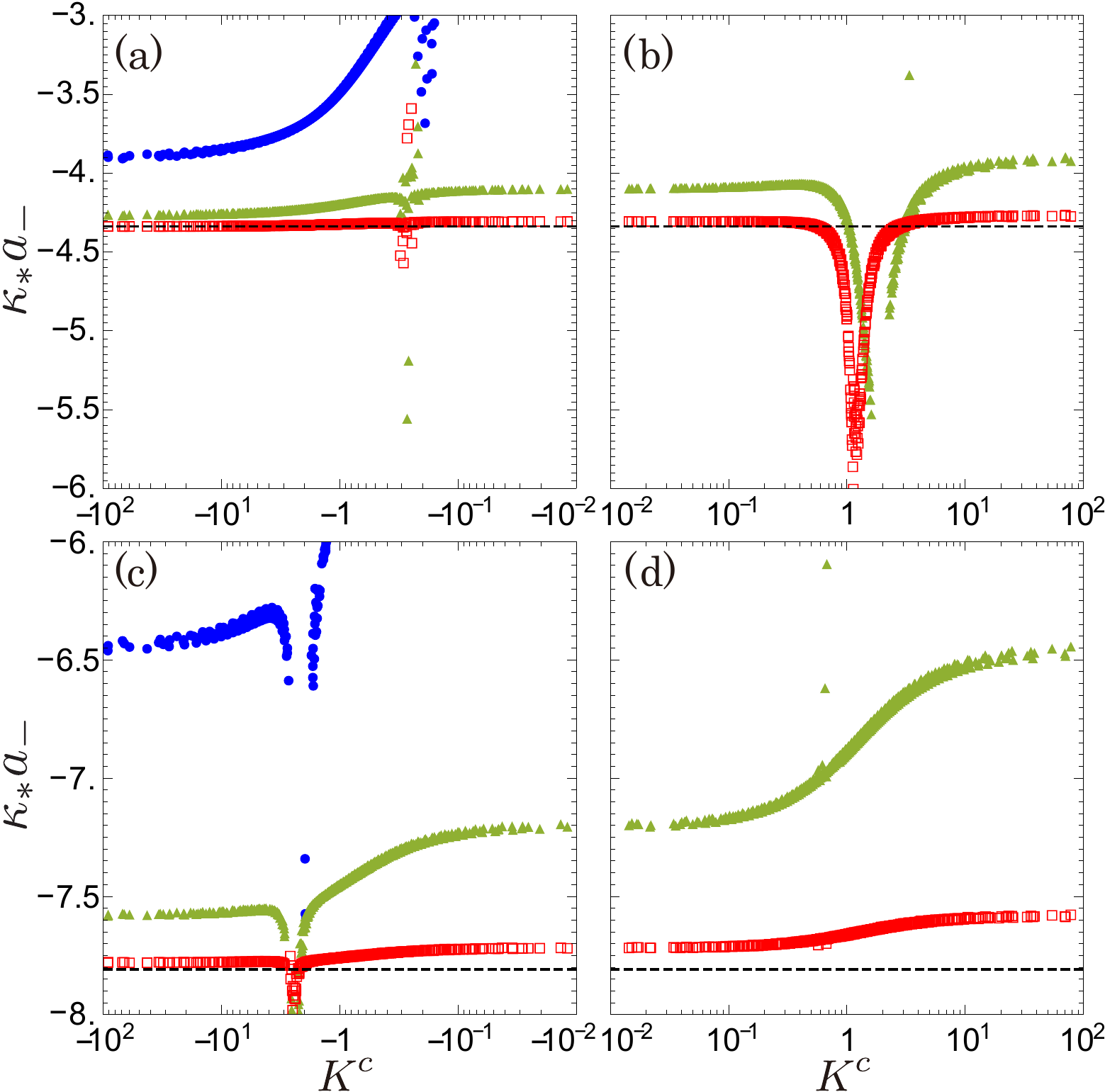} 
\caption{$\kappa_*a_-$ as a function of the quantum defect parameter $K^c$ for (a)(b) the bosonic and (c)(d) fermionic states in $M_z=0$. (a)(b) $a_{\mathrm{dd}}=0.86755...~r_{\mathrm{vdw}}$. (c)(d) $a_{\mathrm{dd}}=0.87054...r_{\mathrm{vdw}}$. The left (right) panels are for negative (positive) $K^c$, which should smoothly connect with each other at $K^c=0$. Blue circles, green triangles, and open red rectangles are the numerical results for the ground, first excited, and second excited Efimov states (see \figref{aHL_Energy}(a)(c)), respectively. Black dashed lines are the universal values of the zero-range Efimov theory, obtained by solving \equref{BOvdwDipoleSchrodinger} with $C_6=C_{\mathrm{dd}}=0$ for a highly excited state.}
\label{fig:Kc_kappaaMinus}
\end{figure}

\subsection{\label{sec:kappatimesaminus}Universality of the Efimov spectrum and $\kappa_*a_- $}
We have tested the universality of the Efimov spectrum by comparing the results obtained with different short-range parameters $R_{\rm{min}}$. While variations in $R_{\rm{min}}$ can generally shift the binding energies of the Efimov states, it can be interpreted as a modification of the three-body parameter. Furthermore, as long as the value of the quantum defect $K^c$ in \equref{QDT_uKc_definition} remains fixed, the Efimov states are expected to exhibit quantitatively similar behavior, hence universal. As demonstrated in Ref~\cite{OiEndo2024} at the unitary limit $1/a^{(\mathrm{HL})}=0$, we have also confirmed that this universality with respect to a change in $R_{\rm{min}}$ holds true for a variable $a^{(\mathrm{HL})}$ in the range $0.25<R_{\rm{min}}/r_{\mathrm{vdw}} <0.40$ we examined, with smaller $R_{\rm{min}}$ showing smaller variance of spectrum originating from difference of $R_{\mathrm{min}}$, hence better universality.

We have also tested the universal relation between the two commonly used three-body parameters: $\kappa_* = \sqrt{M|E|}$, the binding wave number at the unitary limit $1/a^{(\rm{HL})}=0$, and the negative heavy-light scattering length $a_-$ at which the Efimov state dissociates into three atoms. Figure~\ref{fig:Kc_kappaaMinus} shows the dependence of $\kappa_* a_-$ on $R_{\rm{min}}$, reexpressed in terms of the quantum defect parameter $K^c$ using \equref{KcRminrelation2}. In the zero-range low-energy limit of the Efimov theory, these quantities are connected by the universal relation $\kappa_*a_-=\mathrm{const}$, which is denoted by the black-dotted lines in \figref{Kc_kappaaMinus}; $\kappa_*a_- =-4.34$ for bosonic $^{166}$Er-$^6$Li, and $\kappa_*a_- =-7.81$ for fermionic $^{167}$Er-$^6$Li. We have found that the universal relation holds remarkably well for a highly excited state (red open square) for the full range of $K^c$, except for $K^c \simeq 1$ ($K^c \simeq -3$) where an avoided crossing with a higher angular momentum state $L=2$ ($L=3$) breaks the universality for bosons (fermions), respectively (see \figref{aHL_Energy}(b)(d)). In contrast, more deeply bound states show increasing deviation from universality. The first excited state (green triangle) marginally shows the universality with $\sim 10 \%$ accuracy, while the ground Efimov state (blue circle) exhibits substantial deviation. Since cold-atom experiments typically observe Efimov states in the ground or first excited states, which have relatively large binding energies, \figref{Kc_kappaaMinus} suggests that the zero-range universal relation $\kappa_*a_-=\mathrm{const}$ cannot be reliably used to extract $a_-$ from $\kappa_*$ as was done in Ref.~\cite{OiEndo2024}. Instead, accurate quantitative predictions of $a_-$ require three-body calculations with variable $a^{(\rm{HL})}$, as performed in this work, with concrete results presented in \secref{prediction} for several relevant cold-atom isotopes.

\begin{figure}[!t]
	\centering
	\includegraphics[width=1.0\linewidth]{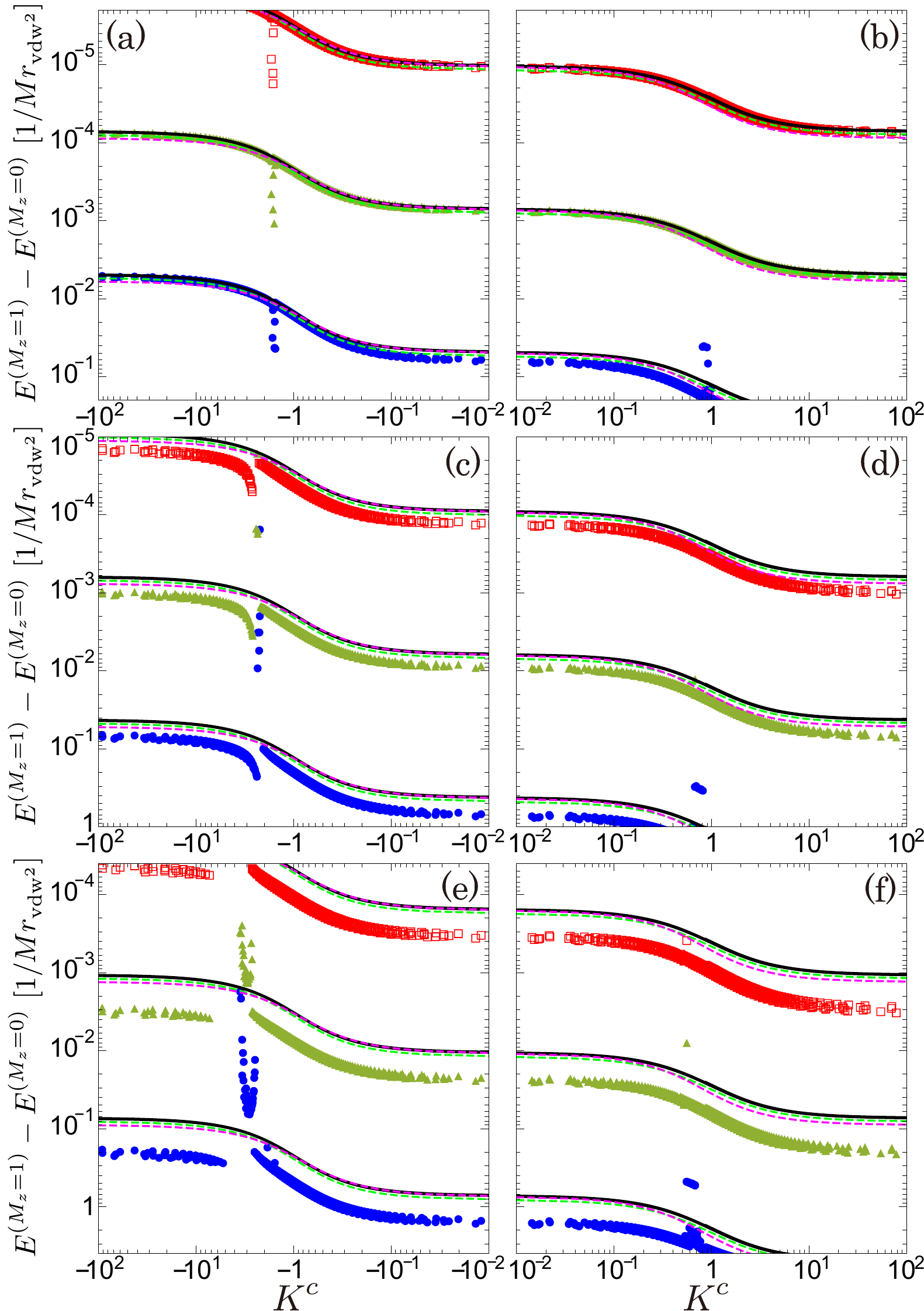} 
\caption{Energy difference between the fermionic Efimov states of $M_z=0$ and $M_z=\pm 1$ at the unitary limit $1/a^{(\rm{HL})}=0$ as a function of the quantum defect parameter $K^c$. Top row: (a)(b) weak dipole regime $a_{\mathrm{dd}}=0.1 r_{\mathrm{vdw}}$. Middle row: (c)(d) moderate dipole regime $a_{\mathrm{dd}}=0.87054...r_{\mathrm{vdw}}$ corresponding to a realistic value for $^{167}$Er-$^6$Li. Bottom row: (e)(f) stronger dipole regime $a_{\mathrm{dd}}=1.5 r_{\mathrm{vdw}}$. The symbols follow the same convention as in \figref{Kc_kappaaMinus}; however, the identification of the ground (blue), first-excited (green), and second-excited (red) states becomes ambiguous near the avoided crossing at $K^c \simeq -3$ and $K^c \simeq 1$, owing to a rearrangement of the energy levels. The black solid, green dashed, and pink dotted curves are the analytical results obtained perturbatively in \equref{split_pert_eq1}, in \equref{split_pert_eq2}, and in \equref{split_pert_eq3}, respectively. }
\label{fig:Kc_EDifMz0Mz1}
\end{figure}

 \subsection{\label{sec:3bodyparameterMz0Mz1}Universality of the three-body parameters between $M_z=0$ and $M_z=\pm1$ states}
In this section, we focus on fermions and their three-body parameters for the two successive Efimov states of $M_z=0$ and $M_z=\pm1$. While the absolute value of the three-body parameter strongly depends on the quantum defect parameter $K^c$~\cite{OiEndo2024}, we investigate here whether the difference between the three-body parameters of $M_z=0$ and $M_z=\pm1$ can exhibit universality. Figure~\ref{fig:Kc_EDifMz0Mz1}~(a)-(f) shows the difference in the three-body binding energy between $M_z=0$ and $M_z\pm1$ states at the unitary limit $1/a^{(\rm{HL})}=0$. The results show a periodic pattern consistent with the discrete scale invariance of the Efimov states: the numerical results appear consistently with the discrete scale invariance of $(e^{\frac{\pi}{|s_1|}})^2=68.2666\dots$ along the vertical axis. Except near the avoided crossings at $K^c \simeq -3$ and $K^c \simeq 1$, the data points with variable $R_{\rm{min}}$ in the range $0.25<R_{\rm{min}}/r_{\mathrm{vdw}} <0.40$ collapse onto universal curves when reexpressed by $K^c$ using \equref{KcRminrelation2}. As the strength of the dipole interaction increases from the top panels [(a)(b)] to the bottom panels [(e)(f)], the energy difference becomes larger, reflecting the lifting of the degeneracy by the introduction of the dipole interaction. These results suggest that the difference between the three-body parameters, and thus the splitting between the three-body loss peaks of $M_z=0$ and $M_z=\pm1$, should be insensitive to $R_{\rm{min}}$, and universally determined by $a_{\mathrm{dd}}$, $r_{\mathrm{vdw}}$ and the quantum defect $K^c$.

\begin{figure*}[!t]
\centering
\begin{minipage}[t]{0.48\linewidth}
\includegraphics[width=1.0\linewidth]{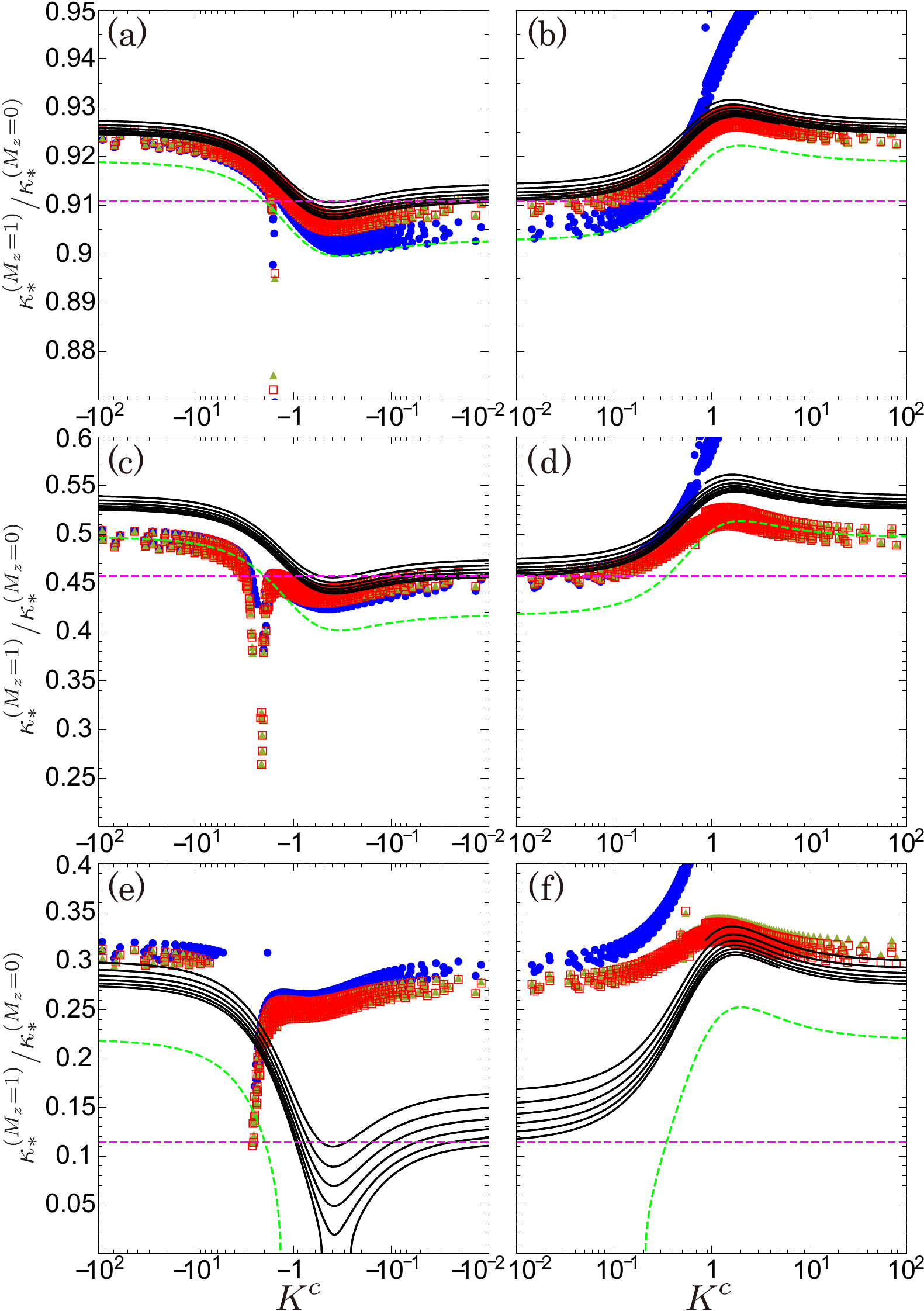} 
 \end{minipage}
 \begin{minipage}[t]{0.48\linewidth}
\includegraphics[width=1.0\linewidth]{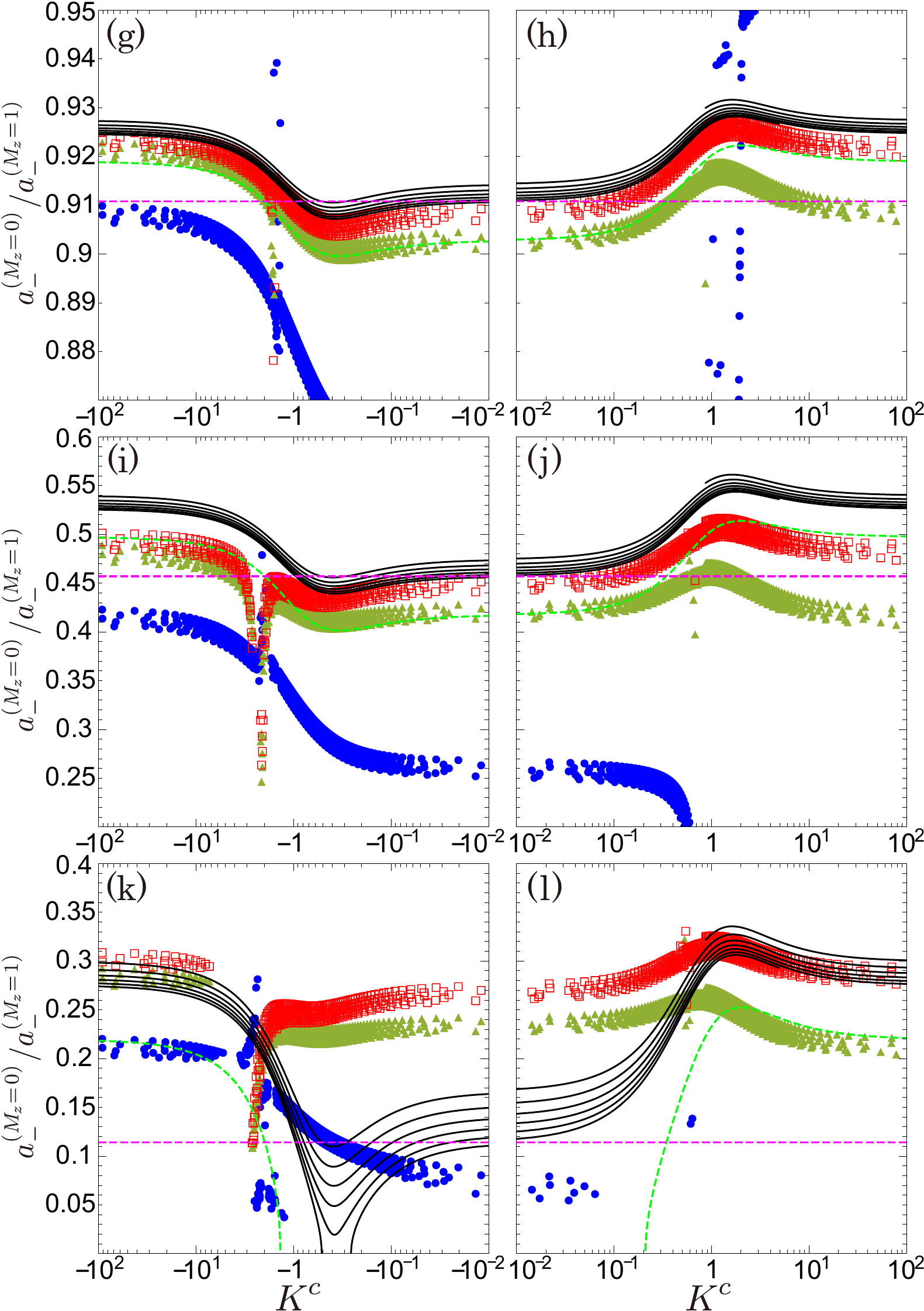} 
 \end{minipage}
\caption{
Ratio of the three-body parameters as a function of the quantum defect parameter $K^c$. Left columns (a)-(f): $\kappa_*^{(M_z=1)}/\kappa_*^{(M_z=0)}$. Right columns (g)-(l) $a_-^{(M_z=0)}/a_- ^{(M_z=1)}$. Top row (a)(b)(g)(h): weak dipole regime $a_{\mathrm{dd}}=0.1 r_{\mathrm{vdw}}$. Middle row (c)(d)(i)(j): moderate dipole regime $a_{\mathrm{dd}}=0.87054...r_{\mathrm{vdw}}$ corresponding to a realistic value for $^{167}$Er-$^6$Li. Bottom row (e)(f)(k)(l): stronger dipole regime $a_{\mathrm{dd}}=1.5 r_{\mathrm{vdw}}$. The symbols are the same as those in \figref{Kc_EDifMz0Mz1}. 
}
\label{fig:Kc_3bodyparamterRatio}
\end{figure*}

To further scrutinize this splitting, we have performed a perturbative analysis. As demonstrated in Ref.~\cite{OiEndo2024}, the energy shifts induced by the dipole interaction follow $\Delta E \propto a_{\mathrm{dd}}$ according to the first-order perturbation theory. By applying it to both $M_z=0$ and $M_z=\pm1$ states, we have obtained the energy splitting at the unitary limit $\Delta E = E^{(M_z =1)}- E^{(M_z =0)}$ as (see Appendix~\ref{app:pertubation})
\begin{equation}
    \begin{split}
        \Delta E  \simeq & \dfrac{18 \pi }{5}|E_0| \dfrac{a_{\mathrm{dd}}}{|s_\ell|\tanh\frac{\pi|s_\ell|}{4}}\left[1+\left(\frac{K^c}{\tanh\frac{\pi|s_\ell|}{4}}\right)^2\right]^{-1}\\
        &\int_{R_{\mathrm{min}}}^\infty dr \frac{\left|A_0\sqrt{r}J_{\frac{s_\ell}{2}}\left(\frac{2r_{\mathrm{vdw}}^2}{r^2}\right)+A_0^*\sqrt{r}J_{-\frac{s_\ell}{2}}\left(\frac{2r_{\mathrm{vdw}}^2}{r^2}\right)\right|^2 }{r^3},
        \label{eq:split_pert_eq1}
    \end{split}
\end{equation}
where $E_0$ is the energy in the absence of the dipole interaction, and $A_0$ is a coefficient of the short-range wave function universally determined by $K^c$ as \equref{A0Kcunivrel}. As shown in Appendix~\ref{app:pertubation}, the integral can be evaluated analytically using the hypergeometric function. Equation~(\ref{eq:split_pert_eq1}) suggests that the ratio $\Delta E / |E_0|$ is independent of the Efimov state's binding energy and thus of its index $n$. In other words, $\Delta E$ exhibits the same discrete scale invariance as $E_0$. This explains why \figref{Kc_EDifMz0Mz1} shows the discrete scale invariant pattern across successive Efimov states. The perturbative formula of \equref{split_pert_eq1} (black solid curves) agrees with the numerical results for weak [(a)(b)] and realistic dipole strengths [(c)(d)], except near the avoided crossings. In contrast, the agreement deteriorates at stronger dipole interactions [(e)(f)] due to the breakdown of the perturbative approximation.

The independence of $\Delta E / |E_0|$ on the Efimov state's index suggests a more universal representation of the results. Figure~\ref{fig:Kc_3bodyparamterRatio}~(a)-(f) [left rows] shows the ratio of the binding wave numbers $\kappa_*^{(M_z=1)}/\kappa_*^{(M_z=0)}$. As can be seen for weak [(a)(b)] and realistic dipole strengths [(c)(d)], the ground, first-excited, and second-excited Efimov states give comparable values, due to the independence of $\Delta E / |E_0|$ on the Efimov state's index. The perturbative expression in \equref{split_pert_eq1} (black solid curves) explains well the overall behavior of the numerical results, except for the ground Efimov state (blue circles) whose relatively large binding energy does not allow the low-energy analysis. The curves are multi-valued for a given $K^c$, reflecting the explicit dependence of \equref{split_pert_eq1} on $R_{\rm{min}}$. The equation can be further simplified by assuming $R_{\rm{min}}/r_{\mathrm{vdw}} \ll 1$, resulting in an $R_{\rm{min}}$ independent expression (see Appendix~\ref{app:pertubation} for the derivation)
\begin{equation}
\begin{split}
 \frac{\Delta E}{|E_0|}  & \simeq  \dfrac{18}{5} \frac{a_{\mathrm{dd}}}{r_{\mathrm{vdw}}}\frac{\sqrt{\pi}\Gamma\left(\frac{1}{4}\right)^2}{8\sqrt{2}|s_\ell|\left|\Gamma\left(\frac{3}{4}+\frac{i|s_\ell|}{2}\right)\right|^2\sinh\frac{\pi|s_\ell|}{2}}\\
 &\left[1+\frac{1}{\sqrt{\cosh(\pi|s_\ell|)}}\cos\left\{\tau^c+\arctan\left(\tanh\frac{\pi|s_\ell|}{2}\right)\right\}\right],
 \label{eq:split_pert_eq2}
 \end{split}
\end{equation}
where $\tau^c/2\equiv \arctan\left(K^c/\tanh\frac{\pi|s_\ell|}{4}\right)$. This universal formula (green dashed curve) agrees well with the numerical results for the first and second excited states in \figref{Kc_3bodyparamterRatio}~(a)-(d), whereas the perturbative analysis breaks down for the ground Efimov state (blue circles), and in the regime of a stronger dipole interaction in \figref{Kc_3bodyparamterRatio}~(e)-(f).

As an experimentally relevant observable for cold atoms, we have also studied the ratio of the three-body parameters $a_-^{(M_z=0)}/a_- ^{(M_z=1)}$, i.e. the positions of the three-body loss peaks, in \figref{Kc_3bodyparamterRatio}~(g)-(l). Due to the universality of $\kappa_*a_- $ demonstrated in \secref{kappatimesaminus}, the ratio $a_-^{(M_z=0)}/a_- ^{(M_z=1)}$ exhibits similar behavior to $\kappa_*^{(M_z=1)}/\kappa_*^{(M_z=0)}$; for the first and second excited states, the numerical results show reasonable agreement with the perturbative predictions obtained from \equref{split_pert_eq1} and \equref{split_pert_eq2} supplemented with the universal value $\kappa_*a_- =-7.81$ obtained in \secref{kappatimesaminus} (black dashed line in \figref{Kc_kappaaMinus}). The results of the ground Efimov state (blue circles) deviate from those of the excited states and from the perturbative curves, reflecting the finite-range effects. While the agreement between the numerical results and the perturbative curves is slightly worse for $a_-^{(M_z=0)}/a_- ^{(M_z=1)}$ than for $\kappa_*^{(M_z=1)}/\kappa_*^{(M_z=0)}$, it is important to note that 
varying $R_\mathrm{min}/r_{\mathrm{vdw}}$ from 0.25 to 0.40 corresponds to a 17-fold change in the depth of the van der Waals interaction. It is thus reasonable to conclude for the Er-Li system [middle panels of \figref{Kc_3bodyparamterRatio}] that the universality of $a_-^{(M_z=0)}/a_- ^{(M_z=1)}$ holds to a limited extent for the ground Efimov state, and to a satisfactory degree for the first and second excited states, which are well captured by the universal formula in \equref{split_pert_eq2}. In contrast, for the Dy-Li system, the dipole interaction is relatively larger $a_{\mathrm{dd}}/r_{\mathrm{vdw}}\simeq 1.6$, corresponding approximately to the bottom panels of \figref{Kc_3bodyparamterRatio}. The perturbative curves can only predict the order of magnitude of $a_-^{(M_z=0)}/a_- ^{(M_z=1)}$ for the Dy-Li system.

While \equref{split_pert_eq2} characterizes the universal behavior of the three-body parameters, it requires knowledge of a quantum defect $K^c$ to make quantitative predictions for $a_-^{(M_z=0)}/a_- ^{(M_z=1)}$. For the bosons, $K^c$ can be evaluated from the $s$-wave scattering length between the heavy atoms, but it is challenging to perform a similar estimate for the fermions~\cite{OiEndo2024}. Notably, this dependence on $K^c$ entirely disappears in the limit of large mass imbalance $M/m \rightarrow \infty$, hence $|s_{\ell}|\rightarrow \infty$. In this limit, the $\tau^c$ dependent oscillatory term in \equref{split_pert_eq2} vanishes as $\cosh \pi |s_{\ell}|\rightarrow \infty$, resulting in a $K^c$-independent ratio
\begin{equation}
\begin{split}
 \frac{\Delta E}{|E_0|}  \simeq \dfrac{18}{5} \frac{a_{\mathrm{dd}}}{r_{\mathrm{vdw}}}\frac{\sqrt{\pi}\Gamma\left(\frac{1}{4}\right)^2}{8\sqrt{2}|s_\ell|\left|\Gamma\left(\frac{3}{4}+\frac{i|s_\ell|}{2}\right)\right|^2\sinh\frac{\pi|s_\ell|}{2}}.
 \label{eq:split_pert_eq3}
 \end{split}
\end{equation}
For Er-Li system shown in \figref{Kc_3bodyparamterRatio}, this universal constant supplemented with the universal value of $\kappa_*a_- =-7.81$ (pink dotted line) agrees reasonably well with the numerical results despite its simplicity. This is because $\sqrt{\cosh (\pi |s_{\ell}|)} =7.31...$ for the Er-Li mass ratio, suggesting that the $K^c$ dependence introduces at most 15\% correction to the universal constant. Therefore, even without any knowledge on the quantum defect $K^c$, we can still make a semi-quantitative prediction solely from the mass ratio and $a_{\mathrm{dd}}/r_{\mathrm{vdw}}$ using \equref{split_pert_eq3} and $\kappa_*a_- =\mathrm{const.}$. 
We note however that \equref{split_pert_eq3} cannot be applied to Dy-Li because the energy in $M_z=\pm1$ states becomes positive in the mass-imbalanced limit Eqs.~(\ref{eq:Energy1storderPerturb}) (\ref{eq:1ovr3ExpValueSmallRminLargeMR}), representing a breakdown of the first-order perturbation. Nevertheless, as shown in Figs.~\ref{fig:Kc_3bodyparamterRatio}(e)(f)(k)(l) (and in \secref{prediction} for Dy-Li system), the numerical results of the ratio of three-body parameter in the excited states mostly lies within a relatively narrow range 0.20--0.35 even in stronger dipole regime, implying that the universal property for a highly mass-imbalanced system persists beyond the perturbative regime. 

In the above discussions on the universality based on the quantum defect $K^c$, the dipole interaction is neglected in the determination of $K^c$ in Eq.~(\ref{eq:KcRminrelation2}). This approximation is justified because the dipole interaction is negligibly small -- at least 50 times weaker than the van der Waals force -- at short distance $r \simeq R_{\mathrm{min}}$ for the range explored in this work. It can be even smaller for a realistic value of $R_{\mathrm{min}}$ (see \secref{prediction}).

 \subsection{\label{sec:prediction}Quantitative Prediction for Er-Li and Dy-Li experiments}
In this section, we provide quantitative predictions for the Efimov states in Er-Li and Dy-Li mixtures. Specifically, we use realistic parameters for the mass ratio, $r_{\mathrm{vdw}}$, and $a_{\mathrm{dd}}$~\cite{PhysRevX.5.041029,Chomaz_2023}.

\begin{table*}[!t]
\caption{Three-body parameters for the bosonic Er-Er-Li and Dy-Dy-Li Efimov states. $\kappa_* =\sqrt{M|E|}$ denotes the binding wave number at unitary limit $1/a^{(\mathrm{HL})}=0$, and $a_-$ is the scattering length $a^{(\mathrm{HL})}$ where the Efimov states dissociate into three atoms. For each species, $\kappa_*$, $a_-$ and $\kappa_*a_-$ are shown from the ground state (top) to the third-excited state (bottom). For $^{166}$Er-$^{6}$Li, $^{160}$Dy-$^{6}$Li and $^{164}$Dy-$^{6}$Li, the upper (lower) set of results are for the upper (lower) side of the resonance in Fig. 5 in Ref.~\cite{OiEndo2024}. Our numerical calculations use the van der Waals and dipole lengths reported in Refs.~\cite{PhysRevX.5.041029,Chomaz_2023} along with the mass scaling. The heavy-heavy scattering length $a^{\mathrm{(HH)}}$ is estimated as the background scattering length $a_{bg}$, whose value is reported in Ref.~\cite{Chomaz_2023}.}
\centering
\label{tab:result_list_boson} 
\begin{tabular}{cccccccc }
\hline
Species & $r_{\mathrm{vdw}} $[$a_0$]& $a_{\mathrm{dd}}$ [$a_0$] & $a^{(\mathrm{HH})}$ [$a_0$] & $\kappa_* r_{\mathrm{vdw}} $ & $a_-/r_{\mathrm{vdw}}$& $\kappa_*a_-$\\
 \hline
 \multirow{8}{*}{$^{166}$Er-$^{6}$Li}& \multirow{8}{*}{75.5} & \multirow{8}{*}{65.5} & \multirow{8}{*}{68} & 0.40\ -\ 0.41 & {-9.6}\ --\ {-9.9} & {-3.9}\ --\ {-4.0} \\
 & & & & 8.7$\times 10^{-2}$ & -49 & -4.3 \\
 & & & & 1.9$\times 10^{-2}$ & {-2.3$\times10^2$}& {-4.3}\\
 & & & & 4.0$\times 10^{-3}$\ --\  4.1$\times 10^{-3}$&  {-1.1$\times 10^{3}$} & {-4.3} \\

 %0.51\ --\ 0.53 & {-7.1}\ --\ {-7.3} & -3.7 \\
% & & & & 0.11 & {-38}\ --\ {-39} & -4.2 \\
% & & & & 2.3$\times 10^{-2}$\ --\  2.4$\times 10^{-2}$ & {-1.8$\times10^2$}\ --\ {-1.9$\times10^2$} & -4.3 \\
% & & & & 5.0$\times 10^{-3}$\ --\  5.2$\times 10^{-3}$&  {-8.4$\times10^2$}\ --\ {-8.7$\times10^2$} & -4.3 \\
 \cline{5-7}
& & & & 0.51\ --\ 0.53 & {-7.1}\ --\ {-7.3} & -3.7 \\
 & & & & 0.11 & {-38}\ --\ {-39} & -4.2 \\
 & & & & 2.3$\times 10^{-2}$\ --\  2.4$\times 10^{-2}$ & {-1.8$\times10^2$}\ --\ {-1.9$\times10^2$} & -4.3 \\
 & & & & 5.0$\times 10^{-3}$\ --\  5.2$\times 10^{-3}$&  {-8.4$\times10^2$}\ --\ {-8.7$\times10^2$} & -4.3 \\
% & & & &  0.40\ -\ 0.41 & {-9.6}\ --\ {-9.9} & {-3.9}\ --\ {-4.0} \\
% & & & & 8.7$\times 10^{-2}$ & -49 & -4.3 \\
% & & & & 1.9$\times 10^{-2}$ & {-2.3$\times10^2$}& {-4.3}\\
% & & & & 4.0$\times 10^{-3}$\ --\  4.1$\times 10^{-3}$&  {-1.1$\times 10^{3}$} & {-4.3} \\
 \hline
 \multirow{4}{*}{$^{168}$Er-$^{6}$Li}& \multirow{4}{*}{75.8} & \multirow{4}{*}{66.3} & \multirow{4}{*}{137} & 0.50\ --\ 0.52&{-10} & {-5.0}\ --\ {-5.3}\\
 & & & & 0.22\ --\ 0.25& {-29}\ --\ {-32} & -7.2 \\
 & & & & 5.6$\times 10^{-2}$\ --\ 6.1$\times 10^{-2}$& {-78}\ --\ {-91} & {-4.8}\ --\ {-5.0} \\
 & & & & 1.2$\times 10^{-2}$\ --\ 1.3$\times10^{-2}$&  {-3.3$\times10^2$}\ --\ {-3.7$\times10^2$} & {-4.4}\ --\ {-4.5} \\
 \hline
 \multirow{4}{*}{$^{170}$Er-$^{6}$Li}& \multirow{4}{*}{76.0} & \multirow{4}{*}{67} & \multirow{4}{*}{221} & 0.32\ --\ 0.33 & -13 & -4.2\\
 & & & & 7.3$\times 10^{-2}$\ --\ 7.4$\times 10^{-2}$& {-60}\ --\ {-62} & -4.5 \\
 & & & & 1.6$\times 10^{-2}$& {-2.7$\times10^2$}\ --\ {-2.8$\times10^2$} & {-4.4} \\
 & & & & 3.5$\times 10^{-3}$\ --\ 3.6$\times 10^{-3}$& {-1.2$\times 10^{3}$}\ --\ {-1.3$\times 10^{3}$} & {-4.4} \\
 \hline
 \multirow{8}{*}{$^{160}$Dy-$^{6}$Li}& \multirow{8}{*}{77.72} & \multirow{8}{*}{127.6} & \multirow{8}{*}{74.3} & 0.44\ --\ 0.45\ & {-8.8}\ --\ {-8.9} & {-3.9}\ --\ {-4.0}\\
 & & & & {0.94$\times 10^{-1}$}\ --\ {1.0$\times10^{-1}$}& {-45}\ --\ {-46} & {-4.3}\ --\ {-4.5} \\
 & & & & 2.0$\times 10^{-2}$\ --\ 2.1$\times 10^{-2}$& {-2.1$\times10^2$}\ --\ {-2.2$\times10^2$} & -4.3 -- -4.4\\
 & & & & 4.1$\times 10^{-3}$\ --\ 4.4$\times 10^{-3}$& {-9.7$\times 10^{2}$}\ --\ {-1.0$\times10^{3}$} & -4.3 \\
 \cline{5-7}
 & & & & 0.59\ --\ 0.63\ & {-5.5}\ --\ {-5.9} & {-3.4}\\
 & & & & {0.12}\ --\ {0.13}& {-31}\ --\ {-34} & {-4.2} \\
 & & & & 2.6$\times 10^{-2}$\ --\ 2.8$\times 10^{-2}$& {-1.5$\times10^2$}\ --\ {-1.6$\times10^2$} & {-4.3} \\
 & & & & 5.4$\times 10^{-3}$\ --\ 5.8$\times 10^{-3}$& {-7.3$\times10^2$}\ --\ {-7.9$\times10^2$} & -4.3 \\
  \hline
 \multirow{4}{*}{$^{162}$Dy-$^{6}$Li}& \multirow{4}{*}{77.97} & \multirow{4}{*}{129.2} & \multirow{4}{*}{157} & 0.32\ --\ 0.33\ & -12 & -3.9\\
 & & & & {6.8$\times10^{-2}$}\ --\ {6.9$\times10^{-2}$}& {-61}\ --\ {-62} & -4.2 \\
 & & & & 1.4$\times 10^{-2}$\ --\ 1.5$\times 10^{-2}$& {-2.9$\times10^2$}\ --\ {-3.0$\times10^2$} & -4.3 \\
 & & & & 3.0$\times 10^{-3}$\ --\ 3.1$\times 10^{-3}$& {-1.4$\times10^3$} & {-4.3} \\
  \hline
 \multirow{8}{*}{$^{164}$Dy-$^{6}$Li}& \multirow{8}{*}{78.21} & \multirow{8}{*}{130.7} & \multirow{8}{*}{92} & 0.41 & {-9.4}\ --\ {-9.6} & {-3.9}\ --\ {-4.0}\\
 & & & & {8.9$\times10^{-2}$}\ --\ {9.0$\times10^{-2}$}& {-48} & {-4.3}\ --\ {-4.4} \\
 & & & & 1.9$\times 10^{-2}$ & {-2.2$\times10^2$}\ --\ {-2.3$\times10^2$} & -4.3 \\
 & & & & 4.0$\times 10^{-3}$\ --\ 4.1$\times 10^{-3}$& {-1.0$\times10^3$}\ --\ {-1.1$\times10^3$} & -4.3 \\
\cline{5-7}
 & & & & 0.61\ --\ 0.64\ & {-5.4}\ --\ {-5.7} & {-3.4}\ --\ {-3.5}\\
 & & & & {0.13}\ --\ {0.14}& {-30}\ --\ {-32} & {-4.2} \\
 & & & & 2.8$\times 10^{-2}$\ --\ 2.9$\times 10^{-2}$& {-1.4$\times10^2$}\ --\ {-1.5$\times10^2$} & -4.3 \\
 & & & & 5.9$\times 10^{-3}$\ --\ 6.4$\times 10^{-3}$& {-6.8$\times10^2$}\ --\ {-7.3$\times10^2$} & -4.3 \\
  \hline
\end{tabular}
\end{table*}

We show the three-body parameters of the bosonic Efimov states $M_z=0$ for various Er and Dy isotopes in \tabref{result_list_boson}. When numerically solving \equref{BOvdwDipoleSchrodinger}, we choose the value of $R_{\mathrm{min}}$ to reproduce the typical background heavy-heavy $s$-wave scattering length $a^{(\mathrm{HH})}$ as listed in \tabref{result_list_boson}. More specifically, we obtain the relation between $a^{(\mathrm{HH})}$ and $R_{\mathrm{min}}$ by solving the Schr\"{o}dinger equation for two heavy atoms interacting via the dipole and van der Waals interactions (i.e., without $V_{\mathrm{BO}}$), from which we identify the values of $R_{\mathrm{min}}/r_{\mathrm{vdw}}$ that reproduces realistic values of $a^{(\mathrm{HH})}/r_{\mathrm{vdw}}$ (see Fig. 6 in Ref.~\cite{OiEndo2024}). The uncertainty in $\kappa_* $ and $a_-$ arises from the ambiguity in $R_{\mathrm{min}}$, since different values of $R_{\mathrm{min}}/r_{\mathrm{vdw}}$ can yield the same $a^{(\mathrm{HH})}$. To account for this, we have performed numerical calculations with all possible values of $R_{\mathrm{min}}/r_{\mathrm{vdw}}$ in the range 0.25--0.40, and determined the error bars from the resulting variation in the three-body parameters. The small error bars reflect the universal nature of the Efimov states with respect to short-range details $R_{\mathrm{min}}$ (see \secref{kappatimesaminus}). For Er-Li, the values of $\kappa_* r_{\mathrm{vdw}}$ are in reasonable agreement with those reported in Ref.~\cite{OiEndo2024} using the analytical formula based on the renormalized van der Waals universality. For $^{166}$Er-$^{6}$Li, $^{160}$Dy-$^{6}$Li, and $^{164}$Dy-$^{6}$Li, we show two sets of results for each due to an ambiguity in $R_{\mathrm{min}}$; the heavy-heavy scattering length of this isotope is $a^{\mathrm{(HH)}}/r_{\mathrm{vdw}}\simeq 1$, which is near the avoided crossing with the $d$-wave dominant state (see $a^{\mathrm{(HH)}}/r_{\mathrm{vdw}}\simeq 1$ of Fig. 5 in Ref.~\cite{OiEndo2024}). Since the question of which side of the resonance the system lies depends sensitively on the nature of the Feshbach resonance used in the experiments, we have investigated both possibilities and tabulated them in \tabref{result_list_boson}.

While $a_-$ was previously estimated using the zero-range universal relation $\kappa_* a_-=\mathrm{const}$ in Ref.~\cite{OiEndo2024}, as we demonstrated in \secref{kappatimesaminus}, this is only valid for highly excited states and breaks down for the ground Efimov state, which is the most accessible in cold-atom experiments. In the second right-most column of \tabref{result_list_boson}, we show $a_-$ obtained without relying on the zero-range universal relation, instead calculated directly from \equref{BOvdwDipoleSchrodinger}. While our results agree well with those in Ref.~\cite{OiEndo2024} for highly excited states, significant deviations appear for the ground Efimov states, underscoring the importance of the finite scattering length calculation in this work.

The right-most column shows $\kappa_*a_-$. For Er-Li, most values agree excellently with the zero-range universal values of -4.3 to -4.4, corresponding to each mass ratio. The agreement is particularly good for shallower bound state, while the ground Efimov state shows a slight deviation. An exception is the first excited state of $^{168}$Er-$^{6}$Li, where the proximity to a shallow $d$-wave dominant bound state leads to a significantly different value. 

We note that the uncertainty range in \tabref{result_list_boson} is based on the variance for the values explored $ 0.25 \le R_{\mathrm{min}}/r_{\mathrm{vdw}} \le 0.40$, which is likely to be an overestimate. In Ref.~\cite{PhysRevA.105.063307}, the $R_{\mathrm{min}}$ parameter between two $^{168}$Er atoms is crudely estimated with the WKB approximation as $R_{\mathrm{min}} \simeq 0.07 r_{\mathrm{vdw}}$. Since the sensitivity of the results to $R_{\mathrm{min}}$ decreases at smaller values, the actual variation may be significantly smaller than suggested in \tabref{result_list_boson}. The uncertainties due to avoided crossings may be circumvented if one can engineer the value of the three-body parameter~\cite{zulli2025universal}, thereby tuning the Efimov states away from higher angular momentum states.

While we have used in \tabref{result_list_boson} the background scattering length reported in Ref.~\cite{Chomaz_2023} as a plausible value of $a^{\mathrm{(HH)}}$, we can improve our estimates with the knowledge of the value and magnetic field dependence of $a^{\mathrm{(HH)}}$ around the Er-Li and Dy-Li Feshbach resonance of interests.

\begin{table}[!t]
\caption{Ratio of the three-body parameters between the $M_z=0$ and $M_z=\pm1$ states for the fermionic Er-Er-Li and Dy-Dy-Li systems. In the second right-most column, the ratio obtained with the perturbative expression in \equref{split_pert_eq3} is shown for Er-Li, while it is not shown due to the breakdown of the perturbation. Range of $a_-^{(M_z=0)}/a_- ^{(M_z=1)}$ evaluated numerically from the entire variation of $K^c$ is shown in the right-most column.}
\centering
\label{tab:result_list_fermion} 
\begin{tabular}{ccccccc}
\hline
Species & $r_{\mathrm{vdw}}$[$a_0$]& $a_{\mathrm{dd}}$[$a_0$] & $\dfrac{a_-^{(M_z=0)}}{a_-^{(M_z=1)}}\Bigg|_{\frac{M}{m}=\infty}$ & $\dfrac{a_-^{(M_z=0)}}{a_- ^{(M_z=1)}}$\\
\hline
  \multirow{3}{*}{$^{167}$Er-$^{6}$Li} &  \multirow{3}{*}{75.7} & \multirow{3}{*}{65.9} &  \multirow{3}{*}{0.46} & $<$0.42 \\ & & & & 0.41--0.49 \\ & & & & 0.43--0.52 \\
  \hline
 \multirow{3}{*}{$^{161}$Dy-$^{6}$Li}& \multirow{3}{*}{77.85} & \multirow{3}{*}{128.4} & \multirow{3}{*}{ } & $<$0.17 \\ & & & & 0.16--0.24 \\ & & & & 0.19-0.28 \\
 \hline
 \multirow{3}{*}{$^{163}$Dy-$^{6}$Li}& \multirow{3}{*}{78.09} & \multirow{3}{*}{130} & \multirow{3}{*}{ } & $<$0.17 \\ & & & & 0.16--0.23 \\ & & & & 0.19-0.28 \\
  \hline
\end{tabular}
\end{table} 

For fermionic systems, it is challenging to make a quantitative prediction of the three-body parameter in the same manner by determining $R_{\mathrm{min}}$ and $K^c$, owing to a weak dependence of the low-energy scattering parameter on $R_{\mathrm{min}}$ in fermions~\cite{bohn2009quasi,OiEndo2024,PhysRevA.64.022717,PhysRevA.78.040703,PhysRevA.85.022704}. Nevertheless, using the results presented in \secref{3bodyparameterMz0Mz1}, we can predict the ratio between the three-body parameters of $M_z = 0$ and $M_z = \pm1$ states.  \tabref{result_list_fermion} shows our estimate of the ratio of the three-body parameters, obtained by using the $K^c$ independent formula in \equref{split_pert_eq3} and the zero-range limit value of $\kappa_*a_-$. As demonstrated in \figref{Kc_3bodyparamterRatio}, the large mass-imbalance-limit formula in \equref{split_pert_eq3} holds with 10-15\% accuracy for $^{167}$Er-$^6$Li. As $\kappa_*a_-$ for the ground Efimov state deviates from the zero-range limit value by $\lesssim$ 20\%, and much less for the excited states, our estimates for Er-Li are expected to be reliable within about 25\% accuracy. In contrast, for Dy-Li, the energy of the $M_z=\pm1$ state becomes positive according to \equref{1ovr3ExpValueSmallRminLargeMR}, indicating that Dy behaves as non-perturbative dipolar atoms. 

We also present in \tabref{result_list_fermion} the possible range of the ratio $a_-^{(M_z=0)}/a_- ^{(M_z=1)}$, evaluated from its variations across the full range $0.25 \le R_{\mathrm{min}}/r_{\mathrm{vdw}} \le 0.40$ (i.e., based on all the data points shown in \figref{Kc_3bodyparamterRatio}(i)(j)). $a_-^{(M_z=0)}/a_- ^{(M_z=1)}$ for the higher excited states is found to lie universally within a narrow range. In contrast, for the ground Efimov state, we can only evaluate the upper bound, as $a_-^{(M_z=0)}/a_- ^{(M_z=1)}$ suddenly decreases as $K^c$ is varied (see blue circles in \figref{Kc_3bodyparamterRatio}(i)(j)). Since this is due to a strong finite-range effect for the ground Efimov state of $ \kappa_* \gtrsim r_{\mathrm{vdw}}^{-1}$ and $|a_-|\lesssim r_{\mathrm{vdw}} $, $a_-^{(M_z=0)}/a_- ^{(M_z=1)}$ should be experimentally observed within the narrow range tabulated for the higher excited states provided the experiment is conducted for $|a|\gg r_{\mathrm{vdw}}$. Specifically, we predict the ratio between the two adjacent three-body loss peaks of the fermionic Efimov states as $ 0.41 \le a_-^{(M_z=0)}/a_- ^{(M_z=1)} \le 0.52$ for $^{167}$Er-$^6$Li, and $ 0.16 \le a_-^{(M_z=0)}/a_- ^{(M_z=1)} \le 0.28$ for $^{161}$Dy-$^6$Li and $^{163}$Dy-$^6$Li. The predicted range is narrow for $^{167}$Er-$^6$Li, as this system lies marginally within the perturbative regime, with \equref{split_pert_eq3} closely approximating the central value. In contrast, the range is broader for Dy-Li due to the stronger dipole interaction, which leads to greater variation with respect to $K^c$ and $R_{\mathrm{min}}$ (see the middle and bottom rows of \figref{Kc_3bodyparamterRatio}).

The physical mechanism of the splitting of the $M_z=0$ and $M_z= \pm 1$ Efimov states is similar to the splitting of $p$-wave Feshbach resonances~\cite{PhysRevA.69.042712,PhysRevA.70.030702,PhysRevA.71.045601,PhysRevA.88.012710,PhysRevA.85.051602,PhysRevA.100.050701,luciuk2016evidence}, which also originates from the finite angular momentum and the dipole interaction. We note however that the ratio $a_-^{(M_z=0)}/a_- ^{(M_z=1)}$ is so large that the splitting in terms of the magnetic field can be much larger than that of the $p$-wave Feshbach resonance doublet. This is partly because the dipole interaction of Er and Dy is much larger than the other cold atoms. Another reason is that our Efimov system does not exhibit a repulsive centrifugal barrier, and therefore is free from a tunneling mechanism responsible for a narrow magnetic field width of the $p$-wave Feshbach resonance and its doublet~\cite{chin2010feshbach}.

Effects neglected in our theoretiacl model may cause discrepancies between our predictions in Tables~\ref{tab:result_list_boson} and~\ref{tab:result_list_fermion} and experimental three-body parameters. First, non-adiabatic corrections beyond the Born-Oppenheimer approximation may slightly modify the three-body parameter~\cite{PhysRevLett.109.243201,PhysRevA.95.062708}, along with the universal discrete scale factor. Second, the finite-range nature of the heavy-light interaction, assumed as a zero-range contact interaction in this work, may have a more significant effect. With the Er-Li van der Waals length being nearly half that of Er-Er, finite-range corrections may be non-negligible particularly for $\kappa_*  \gtrsim 0.5 r_{\mathrm{vdw}}^{-1}$ and $|a_-| \lesssim 10 r_{\mathrm{vdw}}$.

\section{\label{sec:Concl}Conclusion}
We have studied Efimov states in a highly mass-imbalanced three-body system composed of two identical heavy atoms and a light atom using the Born-Oppenheimer approximation. Focusing specifically on the Er-Er-Li and Dy-Dy-Li cold-atoms in the vicinity of a heavy-light broad $s$-wave Feshbach resonance, we have demonstrated that the Efimov spectra are universal over a wide range of the $s$-wave scattering length, even when the dipole interaction is as strong as the van der Waals interactions. In particular, we have shown that the three-body parameters characterizing the Efimov states are universal (i.e., insensitive to short-range details) in both bosonic and fermionic systems. 

Based on this universality, we present quantitative predictions for the three-body parameters $\kappa_*$ and $a_-$ in the Er-Li and Dy-Li systems. In the bosonic system where the Efimov states appear only in $M_z =0$ channel, the three-body parameters can be evaluated from the $s$-wave scattering length between the heavy atoms as summarized in Table~\ref{tab:result_list_boson}. In contrast, for the fermionic case where the Efimov states can appear in both the $M_z=0$ and $M_z= \pm 1 $ states, direct prediction of the absolute values of $\kappa_*$ and $a_-$ remains challenging due to the lack of a straightforward connection between the low-energy scattering parameter and the short-range quantum defect parameter. Nevertheless, we find that the ratio of the three-body parameters between the $M_z = 0$ and $M_z = \pm 1$ states exhibits appreciable universality. In particular, in the large mass-imbalance limit, we demonstrate the universality by deriving the analytical formula describing the universal ratio. We thus predict a universal doublet structure in the Efimov spectrum, whose three-body loss features at $a_-^{(M_z=0)}$ and $a_-^{(M_z=\pm 1)}$ should appear at positions determined by the universal ratio, as listed in Table~\ref{tab:result_list_fermion}. 

Our work contributes to current experimental pursuits to realize and observe the ``rotating" Efimov states of fermions in  Er–Li~\cite{ErLiFR1,ErLiFR2} and Dy–Li~\cite{DyLifeshbach} mixtures. The study of resonantly interacting three-body systems with a non-trivial interplay between isotropic and anisotropic interactions is also relevant for nuclear physics, where the combination of short-range nuclear forces and anisotropic tensor forces between the nucleons plays a key role in the nuclear stability and reactions~\cite{PhysRevLett.95.232502,TANIHATA2013215,AnnRev_HamPlatt,PhysRevLett.120.052502,hammer2017effective,endoepelbaumQcl2024}. Taking advantage of the universality, our work paves the way for quantum simulations of nuclear few-body phenomena using dipolar cold-atom mixtures~\cite{Chomaz_2023,schafer2020tools}.

\begin{acknowledgments}
 We thank L. Happ, P. Naidon, and Y. Takasu for fruitful discussions. This work was supported by JSPS KAKENHI Grant Numbers JP22K03492, JP23H01174, and JP25K00217. KO acknowledges support from Graduate Program on Physics for the Universe (GP-PU) of Tohoku University. SE acknowledges support from Matsuo Foundation, and Institute for Advanced Science, University of Electro-Communications.
\end{acknowledgments}
\vspace{\baselineskip}
\appendix

\section{\label{app:pertubation}Calculation of the universal energy splitting with the first order perturbation}
Using the analytical energy and wavefunction in a non-dipole system at the unitary limit $1/a^{(\mathrm{HL})}=0$ obtained in Ref.~\cite{OiEndo2024}, the three-body binding energy can be evaluated perturbatively. Up to the first order, the binding energy in the presence of the dipole interaction is expressed as 
\begin{equation}  \label{eq:Energy1storderPerturb}
E=
    \begin{cases}
        E_0-\dfrac{12}{5}\dfrac{a_{\mathrm{dd}}}{M}\left\langle\dfrac{1}{r^3}\right\rangle  &   \text{$(M_z=0)$}  \\
        E_0+\dfrac{6}{5}\dfrac{a_{\mathrm{dd}}}{M}\left\langle\dfrac{1}{r^3}\right\rangle        &   \text{$(M_z=\pm 1)$}
    \end{cases}
\end{equation}
where
\begin{equation}
\label{eq:1ovr3ExpValue}
   \left\langle\dfrac{1}{r^3}\right\rangle= \frac{\displaystyle\int_{R_\mathrm{min}}^{\infty}
   \frac{|u_\ell(r)|^2}{r^3}dr}{\displaystyle\int_{R_\mathrm{min}}^{\infty }|u_\ell(r)|^2dr},
\end{equation}
$u_\ell(r)$ is the radial part of the wave function in the non-dipole system. It is not normalized here for convenience, so the denominator is necessary.

In evaluating the numerator, we note that the integral is dominated by the short-range region $r\lesssim r_{\mathrm{vdw}}$. The wavefunction can be approximated by the zero-energy wavefunction~\cite{OiEndo2024,PhysRevA.58.1728,PhysRevA.64.010701}
\begin{equation}
\label{eq:ulshortrangeJs}u_\ell(r)\simeq A_0\sqrt{r}J_{\frac{s_\ell}{2}}(x)+A_0^*\sqrt{r}J_{-\frac{s_\ell}{2}}(x)  \hspace{0.3cm} \text{$(r \lesssim r_{\mathrm{vdw}})$}, 
\end{equation}
where $x = 2 \left(r_{\mathrm{vdw}}/r\right)^2$, and $A_0$ is related with the quantum defect parameter $K^c$ as
\begin{equation}
\label{eq:A0Kcunivrel}
    A_0\equiv \dfrac{1}{2\sqrt{2}\cosh\frac{\pi|s_\ell|}{4}}\left(1-i\dfrac{K^c}{\tanh\frac{\pi|s_\ell|}{4}}\right).
\end{equation}
The numerator in \equref{1ovr3ExpValue} is then evaluated as
\begin{widetext}
\begin{equation}
\label{eq:int1ovrcube}
    \int_{R_{\mathrm{min}}}^\infty \dfrac{|u_\ell(r)|^2}{r^3}dr 
    \simeq \frac{2}{r_{\mathrm{vdw}}}|A_0|^2\int_{R_{\mathrm{min}}/r_{\mathrm{vdw}}}^\infty \frac{J_{\frac{s_\ell}{2}}\left(\frac{2}{\zeta^2}\right)J_{-\frac{s_\ell}{2}}\left(\frac{2}{\zeta^2}\right)}{\zeta^2}d\zeta 
    + \frac{2}{r_{\mathrm{vdw}}}{\mathrm{Re}}\left[A_0^2\int_{R_{\mathrm{min}}/r_{\mathrm{vdw}}}^\infty \frac{J_{\frac{s_\ell}{2}}\left(\frac{2}{\zeta^2}\right)^2}{\zeta^2}d\zeta\right].
\end{equation}
On the other hand, the denominator of \equref{1ovr3ExpValue} is concentrated in the long-range region $r \gg r_{\mathrm{vdw}}$, which can be approximated by its long-range asymptotic form of the modified Bessel function $K$ as~\cite{OiEndo2024,PhysRevA.58.1728,PhysRevA.64.010701}
\begin{equation}  \label{eq:wavefunctionAsymptoticForm}
u_\ell(r)\simeq
        \dfrac{2\sqrt{2}}{\pi}\sinh\frac{\pi |s_\ell|}{4} \sqrt{\cosh\frac{\pi |s_\ell|}{2}}\sqrt{1+\left(K^c/\tanh\frac{\pi |s_\ell|}{4}\right)^2}\sqrt{r}K_{\frac{s_\ell}{2}}(\kappa r)         \hspace{0.3cm}\text{$(r\gg r_{\mathrm{vdw}})$}.
\end{equation}
Here, the prefactor is consistently taken to match the short-range wave function in \equref{ulshortrangeJs}. From this, the denominator in \equref{1ovr3ExpValue} is evaluated as
\begin{equation}
\label{eq:intNorm}
    \begin{split}
        \int_{R_\mathrm{min}}^{\infty }|u_\ell(r)|^2dr 
        \simeq&\frac{8}{\pi^2}\sinh^2\frac{\pi|s_\ell|}{4}\cosh\frac{\pi|s_\ell|}{2}\left[1+\left(K^c/\tanh\frac{\pi|s_\ell|}{4}\right)^2\right]\displaystyle\int_{R_{\mathrm{min}}}^\infty r|K_{i|s_\ell|}(\kappa r)|^2dr  \\
        \simeq& \frac{|s_\ell|\tanh\frac{\pi|s_\ell|}{4}}{\pi\kappa^2}\left[1+\left(K^c/\tanh\frac{\pi|s_\ell|}{4}\right)^2\right],
    \end{split}
\end{equation}
where the final equation is derived using the asymptotic form for $\kappa R_{\mathrm{min}}\ll1$ .

The integrals of a product of the Bessel functions in \equref{int1ovrcube} are given by a hypergeometric function ${}_{2}F_{3}$ as
\begin{align}
\label{eq:intAbsSquaredBessel}
    \int_{R_{\mathrm{min}}/r_{\mathrm{vdw}}}^\infty \frac{J_{\frac{s_\ell}{2}}\left(\frac{2}{\zeta^2}\right)J_{-\frac{s_\ell}{2}}\left(\frac{2}{\zeta^2}\right)}{\zeta^2}d\zeta 
    &=\frac{{}_{2}F_{3} \left( \frac{1}{4}, \frac{1}{2}; \frac{5}{4}, 1-\frac{i|s_\ell|}{2},1+\frac{i|s_\ell|}{2}; -4\left(\frac{r_{\mathrm{vdw}}}{R_{\mathrm{min}}}\right)^4 \right)}{(R_{\mathrm{min}}/r_{\mathrm{vdw}})\left|\Gamma\left(1+\frac{i|s_\ell|}{2}\right)\right|^2}, \\
\label{eq:intSquaredBessel}
    \int_{R_{\mathrm{min}}/r_{\mathrm{vdw}}}^\infty \frac{J_{\frac{s_\ell}{2}}\left(\frac{2}{\zeta^2}\right)^2}{\zeta^2}d\zeta 
    &=\frac{(R_{\mathrm{min}}/r_{\mathrm{vdw}})^{-2i|s_\ell|}{}_{2}F_{3} \left( \frac{1}{4}+\frac{i|s_\ell|}{2}, \frac{1}{2}+\frac{i|s_\ell|}{2}; 1+\frac{i|s_\ell|}{2}, \frac{5}{4}+\frac{i|s_\ell|}{2},1+i|s_\ell|; -4\left(\frac{r_{\mathrm{vdw}}}{R_{\mathrm{min}}}\right)^4 \right)}{(R_{\mathrm{min}}/r_{\mathrm{vdw}})(1+2i|s_\ell|)\Gamma\left(1+\frac{i|s_\ell|}{2}\right)^2}.
\end{align}
For $R_{\mathrm{min}}\ll r_{\mathrm{vdw}}$, Eqs.~(\ref{eq:intAbsSquaredBessel})(\ref{eq:intSquaredBessel}) can be approximated using the asymptotic form of the hypergeometric function as
\begin{align}
\label{eq:intAbsSquaredBesselSmallRmin}
    \int_{R_{\mathrm{min}}/r_{\mathrm{vdw}}}^\infty \frac{J_{\frac{s_\ell}{2}}\left(\frac{2}{\zeta^2}\right)J_{-\frac{s_\ell}{2}}\left(\frac{2}{\zeta^2}\right)}{\zeta^2}d\zeta
    &\simeq
    \frac{\Gamma\left(\frac{1}{4}\right)^2}{4\sqrt{2\pi}\left|\Gamma\left(\frac{3}{4}+\frac{i|s_\ell|}{2}\right)\right|^2}, \\
\label{eq:intSquaredBesselSmallRmin}
    \begin{split}
    \int_{R_{\mathrm{min}}/r_{\mathrm{vdw}}}^\infty \frac{J_{\frac{s_\ell}{2}}\left(\frac{2}{\zeta^2}\right)^2}{\zeta^2}d\zeta 
    &\simeq
     \frac{\Gamma\left(\frac{1}{4}\right)^2}{4\sqrt{2\pi}\left|\Gamma\left(\frac{3}{4}+\frac{i|s_\ell|}{2}\right)\right|^2}\frac{\cosh\frac{\pi|s_\ell|}{2}-i\sinh\frac{\pi|s_\ell|}{2}}{\cosh\pi|s_\ell|}.
     \end{split}
\end{align}
From these, together with \equref{intNorm}, we obtain for $R_{\mathrm{min}}\ll r_{\mathrm{vdw}}$
\begin{equation}
\label{eq:1ovr3ExpValueSmallRmin}
    \begin{split}
        \left\langle\dfrac{1}{r^3}\right\rangle 
        \simeq& \frac{1}{r_{\mathrm{vdw}}}\frac{\sqrt{\pi}\Gamma\left(\frac{1}{4}\right)^2}{8\sqrt{2}|s_\ell|\left|\Gamma\left(\frac{3}{4}+\frac{i|s_\ell|}{2}\right)\right|^2\sinh\frac{\pi|s_\ell|}{2}}\kappa^2 
        \left[1+\frac{1}{\sqrt{\cosh(\pi|s_\ell|)}}\cos\left\{\tau^c+\arctan\left(\tanh\frac{\pi|s_\ell|}{2}\right)\right\}\right],
    \end{split}
\end{equation}
\end{widetext}
where $\tau^c/2\equiv \arctan\left(K^c/\tanh\frac{\pi|s_\ell|}{4}\right)$. This equation consists of two terms: a constant term and an oscillation term with respect to a change in $R_{\mathrm{min}}$. The latter term is suppressed in the large mass-imbalanced limit, namely, $|s_\ell|\gg 1$, so that \equref{1ovr3ExpValueSmallRmin} can be further simplified as 
\begin{equation}
    \label{eq:1ovr3ExpValueSmallRminLargeMR}
     \left\langle\dfrac{1}{r^3}\right\rangle
     \simeq
     \frac{1}{r_{\mathrm{vdw}}}\frac{\sqrt{\pi}\Gamma\left(\frac{1}{4}\right)^2}{8\sqrt{2}|s_\ell|\left|\Gamma\left(\frac{3}{4}+\frac{i|s_\ell|}{2}\right)\right|^2\sinh\frac{\pi|s_\ell|}{2}}\kappa^2.
\end{equation}
Notably, the energy shift in the first order perturbation is proportional to the unperturbed energy $\kappa^2/M$, with its coefficient independent of $K^c$ and $R_{\mathrm{min}}$. Therefore, in the limit of $R_{\mathrm{min}}\ll r_{\mathrm{vdw}}$ and $|s_\ell|\gg 1$, the ratio of the binding wave number at the unitary limit $\kappa_*^{(M_z=0)}/\kappa_*^{(M_z=1)}$ is independent of $K^c$ and $R_{\mathrm{min}}$, and therefore universal.

\end{document}